\documentclass[aps,prb,twocolumn,superscriptaddress]{revtex4-1}

\usepackage{amsmath}
\usepackage{amssymb}
\usepackage{amsfonts}
\usepackage{mathrsfs}
\usepackage{graphicx}
\usepackage{mathrsfs}

\usepackage{color}

\newcommand{\eps}{\varepsilon}
\newcommand{\del}[2]{\frac{\partial #1}{\partial #2}}
\newcommand{\der}[2]{\frac{\text{d} #1}{\text{d} #2}}
\def\r{\mathbf{r}}
\def\A{\mathbf{A}}

\begin{document}

\title{Quantum Brownian motion in a Landau level}

\author{E. Cobanera}
\affiliation{Institute of Theoretical Physics, Center for Extreme Matter and Emergent Phenomena, Utrecht University, Leuvenlaan 4, 3584 CE Utrecht, The Netherlands}
\affiliation{Department of Physics and Astronomy, Dartmouth College, 6127 Wilder Laboratory, Hanover, NH 03755, USA}
\email{Current address. Email: emilio.cobanera@dartmouth.edu}

\author{P. Kristel}
\affiliation{Institute of Theoretical Physics, Center for Extreme Matter and Emergent Phenomena, Utrecht University, Leuvenlaan 4, 3584 CE Utrecht, The Netherlands}

\author{C. Morais Smith}
\affiliation{Institute of Theoretical Physics, Center for Extreme Matter and Emergent Phenomena, Utrecht University, Leuvenlaan 4, 3584 CE Utrecht, The Netherlands}

\date{\today}

\begin{abstract}
Motivated by questions about the open-system dynamics of topological 
quantum matter, we investigated the quantum Brownian motion of an 
electron in a homogeneous magnetic field. When the Fermi length 
\(l_F=\hbar/(v_Fm_{\text{eff}})\) becomes much longer than the magnetic 
length \(l_B=(\hbar c/eB)^{1/2}\), then the spatial coordinates \(X,Y\) 
of the electron cease to commute, \([X,Y]=il_B^2\). As a consequence,
localization of the electron becomes limited by Heisenberg uncertainty, 
and the linear bath-electron coupling becomes unconventional. Moreover,
because the kinetic energy of the electron is quenched by the strong
magnetic field, the electron has no energy to give to or take from the
bath, and so the usual connection between frictional forces and dissipation
no longer holds. These two features make quantum Brownian 
motion topological, in  the regime \(l_F\gg l_B\), which is at the verge of 
current experimental capabilities. We model topological quantum Brownian 
motion in terms of an unconventional operator Langevin equation derived from 
first principles, and solve this equation with the aim of characterizing diffusion.
While diffusion in the noncommutative plane turns out to be conventional, with 
the mean displacement squared being proportional to \(t^\alpha\)and 
\(\alpha=1\), there is an exotic regime for the proportionality constant 
in which it is directly proportional to the friction coefficient 
and inversely proportional to the square of the magnetic field: in this
regime, friction helps diffusion and the magnetic field suppresses all fluctuations. We also show that quantum tunneling can be completely suppressed in the noncommutative plane for suitably designed metastable potential wells, a feature that might be worth exploiting for storage and protection of quantum information.
\end{abstract}

\maketitle


\section{Introduction}
  
While topologically ordered quantum systems\cite{wenbook} remain 
a main theme of current research in condensed matter physics and 
quantum engineering, the focus has been shifting away from 
ground-state and thermal properties, towards the more challenging 
problem of non-equilibrium properties and open-system dynamics. 
In addition to the well developed Floquet theory of classically
driven topological insulators and superconductors, there have been 
investigations of the quench dynamics of prototypical systems 
like the toric code \cite{fazio} and honeycomb\cite{slingerland}
models, and ideas for generalizing the Kibble-Zurek mechanism
to systems without local order parameters.\cite{sondhi} The least 
pursued approach, and the one closest in spirit to this paper, has 
been that of coupling, implicitly by way of the Lindblad equation,
quantum baths to topologically ordered systems,\cite{open_tc}
or designing the non-Hamiltonian part of the Lindblad superoperator
so that it will stabilize known topologically nontrivial 
states.\cite{zoller_majorana} 

Investigations of this latter type are, however, particularly 
intriguing because they point to a far reaching question: What 
is the connection  between the structures associated to
topological ordering in quantum systems, irreversible energy
dissipation, and the theory of quantum open-system dynamics and 
control? In (generalized) gauge theories for example, topological 
ordering is induced by local symmetries,\cite{nussinov} whereas for 
topological insulators and  superconductors, topological ordering 
is induced by global symmetries (chiral, time reversal, and charge 
conjugation) of the single-particle Hamiltonian.\cite{bernevig} From 
this point of view, Refs.~[\onlinecite{open_tc}] and 
[\onlinecite{zoller_majorana}] may be understood as concrete explorations 
of the interplay between these symmetry structures and Lindblad 
(Markovian) open-system dynamics.

In this paper, we investigate the relation between open-system 
dynamics, described within the formalism of the operator Langevin 
equation,\cite{gardiner} and a non-symmetry based source of 
topological ordering: the dimensional reduction of phase space 
experienced by an electron subject to a strong, homogeneous magnetic 
field.\cite{noc} In the limit in which the Fermi length is 
much larger than the magnetic length of the electron, the position 
operators of the electron 
become noncommuting, canonically conjugate variables (with \(\hbar\)
replaced by the magnetic length squared \(l_B^2\)), and its kinetic 
energy becomes quenched (vanishing up to a constant shift). Taken 
together, these two features (a noncommutative configuration space 
and vanishing Hamiltonian) are trademarks of effective topological 
models,\cite{DJT90} and they impact the standard theory\cite{CL83,FLO87} 
of quantum Brownian motion in two ways. First, the coupling of a 
topological model to a bath occurs through noncommuting coordinates, 
and second, because the Hamiltonian vanishes identically in topological 
models, the energetics of the diffusion process are completely controlled 
by the bath. We call this instance of the Brownian-motion problem 
topological. 

Topological quantum Brownian motion displays a surprising mixture of 
conventional and exotic behavior. On one hand, the character of the 
diffusion process is not changed by the passage to the noncommutative 
plane. An electron undergoing normal diffusion (mean squared displacement 
proportional to the time elapsed) before the magnetic field is applied, 
continues to do so after the singular limit of very strong magnetic-field strength is taken. On the other hand, the friction coefficient 
can play an extremely counterintuitive role, since there is a regime 
in which the frictional force helps rather than hinders diffusion! 
Moreover, in this regime the magnetic field suppresses thermal
and quantum fluctuations.

The noncommutative plane is rich in surprises even before considering
Brownian motion. Because the position operators do not commute,
the ability to localize the electron is limited by Heisenberg uncertainty.
In other words, the location of an electron in the noncommutative
plane is always fuzzy. Hence, it seems somewhat paradoxical that,
as we will show, quantum tunneling can be completely suppressed 
in the noncommutative plane. This result immediately suggests
some novel ideas for quantum memories. For example, 
after driving the system into the noncommutative regime 
one could in principle store quantum information stably by 
positioning electrons in some suitably designed potential wells.


The organization of this paper is as follows. In Sec.\,\ref{ncp},
we recall the topological quantum mechanics\cite{DJT90} of an electron 
subject to a strong magnetic field, discuss the physical conditions 
for the emergence of the noncommutative plane, and counterintuitive 
physical aspects like the complete suppression of quantum tunneling.
We conclude by rederiving the noncommutative plane in a different
way, namely by projection onto a Landau level. In Sec.\,\ref{op_leqs} 
we further couple the electron to a bath of electrically neutral,
independent oscillators and derive an operator Langevin equation 
appropriate for modeling topological quantum Brownian motion. 
Finally, in Sec.\,\ref{topobrown} we use our operator Langevin 
equation to investigate a particular aspect of topological 
quantum Brownian motion: diffusion in the noncommutative plane. 
We conclude in Sec.\,\ref{outlook} with a summary and outlook. The 
Appendix, included for completeness, is devoted to calculating 
statistical properties of quantum Langevin forces/velocity fields.

\section{Emergent physics in strong magnetic fields: 
the noncommutative plane}
\label{ncp}

The non-relativistic motion of an electron of mass $m$ and charge $-e$ 
is described by the Lagrangian
\begin{equation}\label{eq:lagrangian}
 L = \frac{1}{2} m \dot{\mathbf{r}}^{2} - 
\frac{e}{c} \dot{\mathbf{r}} \cdot \mathbf{A}(\r) 
+ e\phi(\r)- V(\mathbf{r}), 
\end{equation}
where \(\mathbf{r}\) (\(\dot{\r}\)) is the position (velocity) 
vector of the electron, \(\A\) (\(\phi\)) is the vector (scalar) 
potential describing an external electromagnetic field, \(c\) is 
the speed of light, and $V(\mathbf{r})$ is an external potential. 
Let us focus for concreteness on motion in a homogeneous 
magnetic field applied perpendicularly to the plane of motion. We write 
\(X,Y\) for the Cartesian coordinates in this plane, and adopt the 
Landau gauge $\A = -(BY,0,0)$. Then the Lagrangian function simplifies 
to 
\begin{equation}\label{starting_point}
L = \frac{1}{2} m \left(\dot{X}^{2} + \dot{Y}^{2}\right) 
+ \frac{e}{c} B Y \dot{X} - V(X,Y).
\end{equation}

In the limit of large magnetic field, 
we may neglect the kinetic term, $m(\dot{X}^{2} + \dot{Y}^{2})/2$. 
This approximation will be valid if
\begin{equation}\label{eq:InitialComparison}
\frac{1}{2}m \left( \dot{X}^{2} + \dot{Y}^{2}\right) \ll \frac{e}{c} B Y \dot{X}.
\end{equation}
In solids, \(m=m_{\text{eff}}\) is the effective mass of the electron, 
and the Fermi velocity \(v_F\) provides an upper bound for its 
characteristic velocity. Hence, the left-hand side of 
Eq.~\eqref{eq:InitialComparison} is maximal if $\dot{X}^{2} + \dot{Y}^{2} = v_F^{2}$. Thus, we set $\dot{X} = \dot{Y} = v_{F}/\sqrt{2}$, in which case we obtain
\begin{equation}\label{eq:IntermediateComparison}
    \frac{1}{\sqrt{2}} m_{\text{eff}} v_F \ll \frac{e}{c} B Y.
\end{equation}
If we now set $Y = l_{B} = \sqrt{\hbar c/(eB)}$, we obtain the  
criterion
\begin{equation}
    \frac{1}{\sqrt{2}} m_{\text{eff}} v_F \ll \frac{\hbar}{l_{B}},
\end{equation}
or, equivalently,
\begin{equation}\label{eq:FinalComparison}
    l_{\text{F}} \gg \frac{l_{B}}{\sqrt{2}},
\end{equation}
where $l_{\text{F}} = \hbar/(v_{F} m_{\text{eff}})$ is the Fermi length.

Hence, if the Fermi length $l_{\text{F}}$ is much larger 
than the magnetic length $l_{B}$, then we may use the approximate 
Lagrangian
\begin{equation}\label{eq:ApproximateLagrangian}
\tilde{L} = \frac{e}{c} B Y \dot{X} - V(X,Y),
\end{equation}
for describing the motion of the electron. For the favorable case of electrons in the lowest conduction band of GaAs at room temperature,\cite{blakemore82} 
\begin{eqnarray}
    m_{\text{eff}}=0.063m_{\text{el}},\quad v_F=4.4\times 10^5\text{m}/\text{s} \approx 10^{-3}c.
\end{eqnarray}
Hence, \(l_F=l_B /\sqrt{2} \) for \(B \approx 20\) Tesla.

\subsection{Quantization in ultra-high magnetic fields}
\label{quantihigh}

The general problem of quantizing Lagrangians linear
in the velocity has been discussed in Ref.~[\onlinecite{faddeev88}]. 
For $\tilde{L}$, in particular, there is only one momentum variable 
\begin{equation}\label{eq:canonicalmomentum}
P_{x} = \del{\tilde{L}}{\dot{X}} = \frac{e}{c}BY
\end{equation}
conjugate to \(X\) because $\tilde{L}$ does not depend on $\dot{Y}$. 
The Hamiltonian is then
\begin{equation}
H = P_{x} \dot{X} - \tilde{L} = V(X,Y). 
\end{equation}
Upon quantization, the operators $X$ and $P_{x}$ should obey the 
canonical commutation relations, $[X,P_{x}] = i \hbar $. Hence,
\begin{equation}\label{eq:XYCommutator}
[X,Y]=i\frac{c\hbar}{eB}=il_B^2,
\end{equation}
and this is how the noncommutative planar coordinate emerges in an 
ultra-high magnetic field. 

The noncommutative plane is rich in counterintuitive 
features. Let us explore some of them.

{\it Impossibility of perfect localization.---}
How well, how sharply can we locate an electron moving in the 
noncommutative plane? Since its position operators do not commute,
recall Eq.\,\eqref{eq:XYCommutator}, localization is limited by 
Heisenberg's uncertainty principle. The best one can
do is to prepare the electron in a state that is peaked as sharply
as possible around the mean positions \(\langle X\rangle,\ 
\langle Y \rangle\). These states are the coherent states associated 
with the creation and annihilation operators
\begin{equation}
W=\frac{X+iY}{\sqrt{2}l_B},\quad W^\dagger= \frac{X-iY}{\sqrt{2}l_B}
\end{equation}
(\([W,W^\dagger]=1\)). Let 
\begin{eqnarray}
|w\rangle=e^{wW^\dagger}|0\rangle,
\end{eqnarray}
where \(|0\rangle\) is the unique normalized state satisfying \(W|0\rangle=0\).
Then 
\begin{equation}
\langle w|X|w\rangle= \sqrt{2} l_B\Re(w),\quad 
\langle w|Y|w\rangle=  \sqrt{2}l_B\Im(w),
\end{equation}
where \(\Re(w)\) and \(\Im(w)\) denote the real and the imaginary 
part of \(w\). Moreover, the mean squared dispersion of the 
position of the particle around this average is the minimum allowed 
by Heisenberg's uncertainty principle.  

{\it Complete quenching of quantum tunneling.---} 
Tunneling out of a metastable equilibrium position through 
states that are classically forbidden is a hallmark of quantum 
mechanics, and seems impossible to avoid. Semiclassical
reasoning like the WKB approximation shows it is possible
to suppress the rate of quantum tunneling by applying a 
magnetic field. But what is the precise behavior of the tunneling 
rate in the limit of ultra-high fields? The answer is transparent 
in the noncommutative plane: the rate can converge to zero, so 
that quantum tunneling becomes completely suppressed. 

Consider, for example, the Hamiltonian 
\begin{eqnarray}
H=V(X,Y)=\alpha X^2(1-\beta X)\quad (\alpha,\beta>0)
\end{eqnarray}
in the noncommutative plane, which is independent of the \(Y\) coordinate. 
Normally, an electron positioned anywhere on the metastable 
minimum of this potential well would be able to escape by quantum 
tunneling. Escape is witnessed by the time evolution of 
\(\langle X\rangle \): the electron escapes the well if 
\(\langle X\rangle \) grows beyond a certain value. However, 
in the noncommutative plane, the Hamiltonian of the electron 
is just \(H=V\) and, since \([H,X]=0\), \(\langle X\rangle \)
is constant in time. It follows that the rate of quantum 
tunneling out of this metastable well vanishes in the limit
in which the Fermi length greatly exceeds the magnetic length. 

Another interesting example is provided by the Hamiltonian   
\begin{eqnarray}
H=V(X,Y)=-\alpha(R^2-\beta)^2 \quad (\alpha,\beta>0).
\end{eqnarray}
in the noncommutative plane (inverted Mexican hat).
The observable
\begin{eqnarray}
R^2=X^2+Y^2
\end{eqnarray}
measures the radial distance from the origin, squared.
In order to escape from this potential well, the electron
must be able to change its radial distance to the origin. 
But, since \([H,R^2]=0\), this is not possible, and thus 
we find again that quantum tunneling has been completely 
quenched by the ultra-high magnetic field. 

{\it Rotations and translations.---} 
In the noncommutative plane, \(R^2\) is proportional 
to the infinitesimal generator of rotations. Let 
\begin{eqnarray}
\mathscr{L}=\frac{1}{2l_B^2}R^2. 
\end{eqnarray}
It follows immediately that
\begin{eqnarray}
[\mathscr{L},X]=iY,\quad [\mathscr{L},Y]=-iX.
\end{eqnarray}
Meanwhile, a translation is represented by the unitary 
transformation
\begin{eqnarray}
	U(x_{0},y_{0})=e^{i (y_{0}X- x_{0}Y) /l_B^2},
\end{eqnarray}
since 
\begin{align}
	U(x_{0},y_{0}) X U^\dagger(x_{0},y_{0}) &= X+x_{0},\\
	U(x_{0},y_{0}) Y U^\dagger(x_{0},y_{0}) &= Y+y_{0}.
\end{align}
The relation
\begin{align}
\begin{split}
	U(&x_{0},y_{0}) U(x_{0}',y_{0}') = \\ 
	&= U(x_{0}+x_{0}',y_{0}+y_{0}') e^{i(x_{0}y_{0}'-y_{0}x_{0}')/2l_B^2}
\end{split}
\end{align}
shows that this representation of planar translations is 
projective, as one would expect in the presence of a magnetic field.

\subsection{An alternative point of view: projection onto a Landau level}
\label{alternative}

In this section, we establish the equivalence of projecting 
onto any Landau level and taking the ultra-high-magnetic field 
limit as above.

The canonical momenta corresponding to Eq.~\eqref{eq:lagrangian} are
\begin{equation}
p_{x} = m \dot{X} - \frac{e}{c} A_{x}, \quad
p_{y} = m \dot{Y} - \frac{e}{c} A_{y}.
\end{equation}
In the absence of an electric field and external potential, 
the Hamiltonian corresponding to Eq.~\eqref{eq:lagrangian} 
can be written as
\begin{equation}\label{eq:HallHamiltonian}
H = \frac{1}{2m} \left( \pi_{x}^{2} + \pi_{y}^{2} \right),
\end{equation}
where $\pi_{x}$ and $\pi_{y}$ are the gauge-covariant momenta, 
defined in terms of the usual momenta, 
$p_{x}=-i\hbar\partial/\partial x$ and 
$p_{y}=-i\hbar\partial/\partial y$ as
\begin{equation}
\pi_{x} := p_{x} + \frac{e}{c} A_{x}, 
\quad \pi_{y} := p_{y} + \frac{e}{c} A_{y}.
\end{equation}
The gauge-covariant momenta obey the commutation relation
\begin{equation}\label{eq:InvariantMomentaCommutator}
[\pi_{x},\pi_{y}] = -i \frac{\hbar^{2}}{l_{B}^{2}}.
\end{equation}

The guiding center operators, defined by
\begin{equation}\label{eq:DefGuidingCenter}
C_{x} := X - c\frac{\pi_{y}}{eB}, \quad
C_{y} := Y + c\frac{\pi_{x}}{eB},
\end{equation}
commute with the Hamiltonian \eqref{eq:HallHamiltonian} 
and obey the commutation relation
\begin{equation}\label{eq:GuidingCenterCommutator}
[C_{x},C_{y}] = i l_{B}^{2}.
\end{equation}
Let us now define the Landau-level annihilation and creation 
operators $a$ and $a^{\dagger}$, and the radial annihilation 
and creation operators $b$ and $b^{\dagger}$, respectively, by
\begin{align}
a &= \frac{l_{B}}{\hbar \sqrt{2}}( \pi_{x} - i \pi_{y}), & 
a^{\dagger} &= \frac{l_{B}}{\hbar \sqrt{2}}( \pi_{x} + i \pi_{y}), 
\label{eq:DefLandauLadder} \\
b &= \frac{1}{l_{B} \sqrt{2}}(C_{x} + iC_{y}), & 
b^{\dagger} &= \frac{1}{l_{B} \sqrt{2}}(C_{x} - i C_{y}). 
\label{eq:DefRadialLadder}
\end{align}
The commutation relations
\begin{equation}
[a,a^{\dagger}] = 1, \quad [b,b^{\dagger}] = 1,
\end{equation}
are a straightforward consequence of 
Eqs.~\eqref{eq:InvariantMomentaCommutator} and 
\eqref{eq:GuidingCenterCommutator}. 
Moreover,
\begin{eqnarray}
[a,b]=0=[a,b^\dagger].
\end{eqnarray}

The Hamiltonian Eq.\,\eqref{eq:HallHamiltonian} can 
now be written as
\begin{equation}
H = \hbar \omega_{c}\left(a^{\dagger} a + \frac{1}{2}\right),
\end{equation}
where $\omega_{c} = eB/(mc)$ is the cyclotron frequency. 
The identity
\begin{equation}
2l_{B}^{2}\left( b^{\dagger}b - \frac{1}{2} \right) = C_{x}^{2} + C_{y}^{2},
\end{equation}
explains why the operators $b$ and $b^{\dagger}$ are called radial.
Since the $a$ and $a^{\dagger}$ commute with $b$ and $b^{\dagger}$, 
a complete basis of normalized eigenvectors of the Hamiltonian 
of Eq.\,\eqref{eq:HallHamiltonian} is constructed as
\begin{equation}
|n,m \rangle = 
\frac{(a^{\dagger})^{n}}{\sqrt{n!}} \frac{(b^{\dagger})^{m}}{\sqrt{m!}} 
|0,0 \rangle,
\end{equation}
where $|0,0 \rangle$ is the unique normalized vector obeying
\begin{equation}
	a |0,0 \rangle = b|0,0 \rangle = 0.
\end{equation}
Hence,
\begin{equation}\label{eq:LandauLevelEnergy}
H |n,m \rangle = 
\hbar \omega_{c}\left( n + \frac{1}{2} \right) |n,m \rangle.
\end{equation}

The fact that the energy does not depend on $m$ indicates that 
$n$ labels the infinitely degenerate Landau levels and $m$ labels 
this degeneracy. From Eq.~\eqref{eq:LandauLevelEnergy}, we see 
that the energy spacing between adjacent Landau levels is
$\hbar \omega_{c} = \hbar e B /(mc)$, i.e.~it is linear in the 
magnetic field. In the large magnetic field limit, we thus see 
that tunneling between Landau levels is suppressed. This explains, 
intuitively, why the large magnetic field limit is equivalent to 
projecting onto a Landau level. Let us make this intuition rigorous. 

The operator that projects a state onto the $n$-th Landau level 
is given by
\begin{equation}
P_{n} = \sum_{m=0}^{\infty} | n, m \rangle \langle n, m|.
\end{equation}
It follows that 
\begin{equation}\label{eq:ProjectorRelation}
	P_{n} a P_{n} = 0, \quad [P_{n},b] = 0.
\end{equation}
Combining Eqs.~\eqref{eq:DefGuidingCenter} and 
\eqref{eq:DefLandauLadder}, we see that
\begin{align}
X &= C_{x} + i \frac{l_{B}}{\sqrt{2}}(a - a^{\dagger}), \\
Y &= C_{y} - \frac{l_{B}}{\sqrt{2}}(a + a^{\dagger}),
\end{align}
which may be combined with Eqs.~\eqref{eq:DefRadialLadder} 
and \eqref{eq:ProjectorRelation} to obtain
\begin{align}
P_{n}XP_{n} &= C_{x}P_{n} = P_{n}C_{x}, \\
P_{n}YP_{n} &= C_{y}P_{n} = P_{n}C_{y}.
\end{align}
Now, one may verify that
\begin{equation}
[P_{n}XP_{n}, P_{n}YP_{n}] = i l_{B}^{2} P_{n},
\end{equation}
which is isomorphic to Eq.~\eqref{eq:XYCommutator} on the 
range of \(P_n\). 

Let us now include an external potential in the Hamiltonian 
\eqref{eq:HallHamiltonian}, and project the Hamiltonian
\begin{equation}
    H = \frac{1}{2m}(\pi_{x}^{2} + \pi_{y}^{2}) + V(X,Y),
\end{equation}
onto the $n$-th Landau level,
\begin{equation}
P_{n}HP_{n} = \hbar \omega_{c} \left( n + \frac{1}{2} \right) 
+ P_{n}V(X,Y) P_{n}.
\end{equation}
If we make the approximation
\begin{eqnarray}
P_nV(X,Y)P_n\approx V(P_nXP_n,P_nYP_n)
\end{eqnarray}
of further neglecting any virtual transitions to other levels, 
then the resulting projected approximate Hamiltonian describes 
precisely the dynamical problem associated to the approximate 
Lagrangian of Eq.~\eqref{eq:ApproximateLagrangian}. It is in 
this sense that this paper investigates Brownian motion restricted 
to a Landau Level.

\section{The Langevin equation in the noncommutative plane}
\label{op_leqs}

In classical mechanics, Brownian motion of a charged 
particle is described by the Langevin equation 
\begin{eqnarray}\label{eq:EMLangevin}
    m\ddot{\r}= -\eta \dot{\r} - e \mathbf{E} - \frac{e}{c}\dot{\r}\times \mathbf{B}
-\nabla V+\mathbf{f},
\end{eqnarray}
where $\eta$ is the friction constant, and where \(\mathbf{f}\) is the random Langevin force satisfying
\begin{eqnarray}
\langle\langle \mathbf{f}\rangle\rangle&=&0\\
\langle\langle f^\alpha(t)f^\beta(t')\rangle\rangle&=&
2\eta k_BT\delta^{\alpha\,\beta}\delta(t-t')\nonumber.
\end{eqnarray}
The operation \(\langle\langle \cdot\rangle\rangle\) denotes
averaging with respect to the probability distribution 
of the random force. 

The classical Langevin equation can be derived from 
a model where the charged particle is coupled to a 
bath of electrically neutral, independent harmonic 
oscillators. A Lagrangian representation of this model is 
\begin{align}\label{eq:BrownianLagrangian}
\begin{split}
L_{\sf Brownian} = \frac{1}{2}& m \dot{\mathbf{r}}^{2} - 
\frac{e}{c} \dot{\mathbf{r}} \cdot \mathbf{A}(\r) + 
e \phi(\r)- V(\mathbf{r})
\\
&+\sum_j\frac{m_j}{2}[\dot{\mathbf{x}}_j^2- \omega_j^2(\mathbf{x}_j-\r)^2],
\end{split}
\end{align}
where the harmonic oscillators have coordinates 
$\mathbf{x}_{j}$, masses \(m_j\), and frequencies 
$\omega_j$. Now suppose that the oscillators 
are in a thermal Gibbs state. Then, the Lagrangian 
\eqref{eq:BrownianLagrangian} can indeed be used
as the starting point for deriving Eq.~\eqref{eq:EMLangevin}, 
see for example appendix C in Ref.~[\onlinecite{caldeira2}]. 

This result is remarkable because the Lagrangian 
\(L_{\sf Brownian}\) can be quantized, and the procedure 
for deriving the classical Langevin equation can be adapted 
in order to derive an operator Langevin equation.\cite{FLO90} 
This standard operator Langevin equation is by now textbook 
material.\cite{gardiner} Nonetheless, we will 
briefly recall its derivation (in one space dimension for 
simplicity) in Sec.\,\ref{standardole} in order to clarify 
some delicate mathematical and physical points and make 
certain ideas and notations readily available for the rest 
of the paper.

Then, in Sec.\,\ref{topolangevin} we will follow the 
same procedure, but starting from the approximation 
\begin{eqnarray}\label{approxLbath}
\tilde{L}_{\sf Brownian} =  
- \frac{e}{c} \dot{\mathbf{r}} \cdot \mathbf{A}(\r) + 
e\phi(\r)- V(\mathbf{r})+\\
\sum_j\frac{m_j}{2}[\dot{\mathbf{x}}_j^2-
\omega_j^2(\mathbf{x}_j-\r)^2],\nonumber
\end{eqnarray}
appropriate for describing motion in an ultra-high magnetic 
magnetic field. Within this approximation the bath is coupled 
to the electron by way of two noncommuting observables. 
This is how, building on a sound foundation, we arrive to 
a Langevin equation for the noncommutative plane.

\subsection{The standard operator Langevin equation}
\label{standardole}

Let us focus on one-dimensional motion, for clarity of
presentation. The starting point is the Hamiltonian
\begin{eqnarray}
H=\frac{P^2}{2M}+V(X)+
\frac{1}{2}\sum_j\left[\frac{p^{2}_{j}}{m_j}+m_j\omega_j^2(x_j-X)^2\right]\ \ \ \
\end{eqnarray}
for a particle with momentum and position operators \(P\) 
and \(X\) respectively, singled out for observation and coupled to a 
bath of independent oscillators labeled by \(j\). 
In the following, we will write
\(\hat{\mathcal{O}}\) for observables in the Schr\"odinger
picture, \(\mathcal{O}(t)=e^{iHt}\hat{\mathcal{O}}e^{-iHt}\)
for the Heisenberg picture, and \(\bar{\mathcal{O}}(t)=
e^{iH_Bt}\hat{\mathcal{O}}e^{-iH_Bt}\) for the interaction
picture, with 
\begin{eqnarray}
    H_B=\frac{1}{2}\sum_j\left[\frac{p^{2}_{j}}{m_j}+m_j\omega_j^2x_j^2\right].
\end{eqnarray}
All three pictures coincide at \(t=0\). 

In order to derive the operator Langevin equation,
we will begin by investigating the dynamics of this closed 
quantum system in the Heisenberg picture. Hence, the state \(\rho\) 
of the system, a density matrix acting on the total Hilbert space
\begin{eqnarray}
\mathcal{H}=\bigotimes_j\mathcal{H}_{x_j}\otimes\mathcal{H}_X
\end{eqnarray}
is independent of time. As a consequence, the presence or absence 
of entanglement between parts of the system at any time other
than \(t=0\) is not directly encoded in \(\rho\). 
The Heisenberg equations of motion are
\begin{eqnarray}
\ddot{x}_j&=&-\omega_j^2(x_j-X),\label{xjmoves}\\
M\ddot{X}&=&-V'(X)+\sum_jm_j\omega_j^2(x_j-X).\label{Xmoves}
\end{eqnarray}

Let us solve the set of equations in the first line. 
Since we would like to use standard results available for
differential equations involving functions, it is safest to
start by solving the associated differential equations for
transition amplitudes. If \(|\Phi\rangle,|\Psi\rangle\)
are normalizable, time-independent states, then
\begin{eqnarray}
\frac{d^2}{dt^2}\langle\Phi|x_j(t)|\Psi\rangle=
-\omega_j^2(\langle\Phi|x_j(t)|\Psi\rangle-\langle\Phi|X(t)|\Psi\rangle).
\end{eqnarray}
The delicate point is whether these matrix elements define 
well-behaved functions of time for which standard manipulations hold. 
The quick answer is yes, \emph{as long as the number of oscillators 
making up the bath is finite}. 

Hence, let us proceed for now under the assumption that 
this is the case. Then, the solution
\begin{align}
    \langle \Phi|x_j(t)|\Psi\rangle&=
\label{sinform}\\
\langle \Phi|\bar{x}_j^h|\Psi\rangle
&+\int_{0}^tds\, \omega_j\sin[\omega_j(t-s)]
\langle \Phi|X(s)|\Psi\rangle, \nonumber
\end{align}
with
\begin{eqnarray}
\bar{x}_j^h(t)=\hat{x}_j\cos(\omega_jt)+
\frac{\hat{p}_j}{m_j\omega_j}\sin(\omega_jt).
\end{eqnarray}
properly incorporates the boundary condition that the 
Heisenberg and Schr\"odinger picture should coincide 
at \(t=0\). Integrating by parts, we obtain the alternative 
representation
\begin{eqnarray}
&&\langle \Phi|x_j(t)|\Psi\rangle-\langle\Phi|X(t)|\Psi\rangle=
\langle \Phi|\bar{x}_j^h|\Psi\rangle
\label{cnumsol}\\
&&-\langle\Phi|\hat{X}|\Psi\rangle\cos(\omega_jt)-
\int_{0}^tds\,\cos[\omega_j(t-s)]
\langle \Phi|\dot{X}(s)|\Psi\rangle. \nonumber
\end{eqnarray}
Notice that the operator \(\bar{x}^h_j(t)\), a solution 
of the homogeneous version of Eq.\,\eqref{xjmoves}, evolves
in time according to the interaction picture, as defined
at the beginning of this section. 

For a finite bath, these time-dependent matrix elements
are reasonably well behaved in general. Hence, we can promote 
the family of c-number solutions of Eq.\,\eqref{cnumsol} to 
operator status,
\begin{eqnarray}\label{bonafide}
&&x_j(t)-X(t)=\bar{x}_j^h(t)\\
&&-\cos(\omega_jt)\hat{X}-\int_{0}^tds\,\cos[\omega_j(t-s)] \dot{X}(s).
\nonumber
\end{eqnarray}
For an infinite bath, this step is ungranted: The many-body 
amplitude \(\langle \Phi|\dot{X}(s)|\Psi\rangle\) will require 
renormalization in general. We will come back to this point
near the end of this section.  

The next step is to substitute Eq.\,\eqref{bonafide}
in Eq.\,\eqref{Xmoves}. In terms of the definitions
\begin{eqnarray}
\mu(t)&=&\sum_jm_j\omega_j^2\cos(\omega_jt),\\
\bar{F}(t)&=&\sum_jm_j\omega_j^2\bar{x}^h_j(t),
\end{eqnarray} 
one obtains
\begin{eqnarray}\label{standardql}
&&M\ddot{X}(t)+V'(X(t))=\\
&&-\int_{0}^{t} ds\,\mu(t-s)\dot{X}(s)-\mu(t)\hat{X}+\bar{F}(t). \nonumber
\end{eqnarray}
Hence, \(\mu\) is the memory kernel. It also appears in the 
commutator
\begin{eqnarray}\label{comf}
[\bar{F}(t),\bar{F}(s)]=i\hbar\frac{d}{dt}\mu(t-s).
\end{eqnarray} 

Equation\,\eqref{standardql} is the standard operator Langevin equation
and the foundation of the Ford-Kac-Mazur approach\cite{FKM65,FLO87,gardiner} 
to modeling dissipation in quantum mechanics. It is a peculiar 
equation of motion because it mixes together the Heisenberg, 
Schr\"odinger, and interaction pictures. In particular,
the quantum Langevin force \(\bar{F}\) evolves in time
according to the interaction picture in which the electron
acts as a perturbation on the bath.

Let us take a closer look at Eq.\,\eqref{standardql},
as it is used for modeling diffusion.\cite{gardiner} Hence, we 
set \(V=0\), and
\begin{eqnarray}\label{inst}
\mu(t)=2\eta\delta(t),
\end{eqnarray} 
so that 
\begin{eqnarray}\label{standarddiff}
M\ddot{X}(t)=-\eta\dot{X}(t)-2\eta\delta(t)\hat{X}+\bar{F}(t),
\end{eqnarray}
according to the rule \(\int^t ds\, \delta (t-s)f(s)=f(t)/2\).
The memory kernel of Eq.\,\eqref{inst} can only be obtained
by letting the number of oscillators in the bath become
infinite, and the term \(-2\eta\delta(t)\hat{X}\) is typical
of the problems associated to this limiting procedure. Taken at 
face value, it indicates that \(P(t)\) is discontinuous at \(t=0\)
(assuming that \(X(t)\) is continuous at \(t=0\)), and so we need 
to be more careful in specifying boundary conditions. If we 
impose 
\begin{eqnarray}\label{crazy}
\lim_{t\rightarrow 0+}P(t)=\hat{P},
\end{eqnarray}
then the solution of the operator Langevin equation
for \(t\geq0\) is 
\begin{eqnarray}\label{toint}
P(t)&=&e^{-\eta t/M}\hat{P}+\int_0^tds\, e^{-\eta(t-s)/M}\bar{F}(s),\\
X(t)&=&
\hat{X}+\frac{1}{\eta}(1-e^{-\eta t/M})\hat{P}\\
&+&\frac{1}{M}\int_0^tds''\int_0^{s''}ds'\,e^{-\eta(s''-s')/M}\bar{F}(s'). \nonumber
\end{eqnarray}

Using these explicit solutions and Eq.\,\eqref{comf} one can show that
\begin{eqnarray}\label{comnice}
[X(t),P(t)]=i\hbar+i\hbar\, e^{-\eta t/M}(e^{-\eta t/M}-1)\quad (t\geq 0).
\ \ \ \ 
\end{eqnarray}
Notice that this commutator is canonical precisely at \(t=0\), 
and for long times, up to exponentially small errors. The 
transient period during which the deviation from \(i\hbar\) is 
appreciable is very short. This result indicates quite correctly 
that making the bath infinite spoils, to some extent, the simple 
microscopic derivation of the operator Langevin equation. As we 
mentioned already, the passage from Eq.\,\eqref{cnumsol} to 
Eq.\,\eqref{bonafide} is not well justified if the bath is 
infinite. In spite of this complication, Eq.\,\eqref{comnice} 
reassures us that the operator Langevin equation remains a
phenomenologically sound starting point for describing quantum 
Brownian motion at times \(t\gg 0\). We would like to stress that the discontinuity at $t=0$ and the partial failure of $[X(t),P(t)] = i \hbar$ are two \emph{separate} subtleties that one has to deal with in this approach. Let us make a final remark on the first of these issues. Within the influence functional formalism advanced by Caldeira and Leggett\cite{caldeira2}, the same kind of problem arises when the bath and the system of interest are considered to be decoupled at $t=0$, (or, equivalently, when the initial state is factorizable). In this case, one also has to perform certain integrals from $t=0^{+}$. A less artificial way to resolve the problem is to consider a state where the system and the bath are initially in thermal equilibrium, see Refs.~[\onlinecite{MC87}] and [\onlinecite{MC90}].

\subsection{The operator Langevin equations in the 
noncommutative regime \(l_C\gg l_B\)}
\label{topolangevin}

In this section, we will go through the steps of deriving 
an operator Langevin equation starting from the 
approximate Lagrangian \(\tilde{L}_{\sf Brownian}\) of 
Eq.\,\eqref{approxLbath}. The associated Hamiltonian is
\begin{align}
\label{noncomelpbath}
\begin{split}
H= &V(X,Y)\\
+&\sum_j \left[ \frac{1}{2m_j}p_j^{x\, 2}+
\frac{1}{2}m_{j} \omega_{j}^{x \, 2}(x_j-X)^2 \right]\\
+&\sum_j \left[ \frac{1}{2m_j}p_j^{y\, 2}+
\frac{1}{2}m_{j} \omega_{j}^{y \, 2}(y_j-Y)^2 \right]
\end{split}
\end{align}
The Heisenberg equations of motion for the oscillators are 
\begin{align}
\ddot{x}_j+\omega_j^{x\, 2}x_j &= \omega_j^{x\, 2}X,\\
\ddot{y}_j+\omega_j^{y\, 2}y_j &= \omega_j^{y\, 2}Y;
\end{align}
and the equations of motion for the electron are
\begin{align}\label{eq:langevinXraw}
	\dot{X} &= \frac{i}{\hbar}[V,X] + \frac{l_B^2}{\hbar}\sum_j m_{j} \omega_{j}^{y \, 2}(y_j-Y), \\
    \dot{Y} &= \frac{i}{\hbar}[V,Y] - \frac{l_{B}^{2}}{\hbar} \sum_{j} m_{j} \omega_{j}^{x \, 2}(x_{j} - X). \label{eq:langevinYraw}
\end{align}
The idea is to reduce this system of equations by solving 
the equations of motion for the oscillators. Let us introduce some notation before we proceed. First, 
\begin{eqnarray}
\label{eq:randomvelocity}
\bar{U}_\alpha(t)=\frac{l_B^2}{\hbar}\sum_j 
m_{j} \omega_{j}^{\alpha \, 2} \bar{r}^\alpha_j(t), 
\end{eqnarray}
with 
\begin{eqnarray}
\bar{r}^\alpha_j(t)=
\hat{r}_j^\alpha \cos(\omega^\alpha_jt) 
+ \hat{p}_j^\alpha \frac{\sin(\omega_j^\alpha t)}{m_j^\alpha\omega_j^\alpha}
\end{eqnarray}
and
\begin{eqnarray}
\hat{r}^x_j=\hat{x}_j,\quad \hat{r}^y_j=\hat{y}_j,
\end{eqnarray} 
defines a random velocity field. The operators 
\(\hat{x}_j,\hat{p}_j^x, \hat{y}_j,\hat{p}_j^y\) are standard, 
time-independent Schr\"odinger position and momentum operators
for the oscillators in the bath. As a consequence, this random 
velocity field evolves in time according to the interaction 
picture. Second,
\begin{eqnarray}
\Omega_\alpha=\frac{l_B^2}{\hbar}\sum_j m_{j} \omega_{j}^{\alpha \, 2} \qquad (\alpha=x,y),
\end{eqnarray}
is a quantity with the dimension of angular frequency, 
and finally
\begin{eqnarray}
\nu_\alpha(t)=\frac{l_B^2}{\hbar}\sum_j
m_{j} \omega_{j}^{\alpha \, 3} \sin(\omega^\alpha_jt)\qquad (\alpha=x,y)
\end{eqnarray}
is a function of time with the dimension of angular 
frequency squared.

In terms of the explicit closed-form expression for \(y_j(t)\), i.e.~the operator $y$ analog of Eq.~\eqref{sinform}, we can 
rewrite Eq.~\eqref{eq:langevinXraw} as
\begin{eqnarray}\label{langevinX}
&&\dot{X}(t)-\frac{i}{\hbar}[V,X](t)=\\
&&\bar{U}_y(t) - \Omega_y Y(t) + \int_0^tds\, \nu_y(t-s)Y(s).\nonumber
\end{eqnarray}
Similarly,
\begin{eqnarray}\label{langevinY}
&&\dot{Y}(t)-\frac{i}{\hbar}[V,Y]=\\
&&-\bar{U}_x(t) + \Omega_x X(t)- \int_0^t ds\, \nu_x(t-s)X(s). \nonumber 
\end{eqnarray}
Moreover, in terms of standard memory kernels
\begin{equation}\label{defmemker}
\mu_{\alpha}(t) = \frac{l_{B}^{2}}{\hbar} 
\sum_{j} m_{j} \omega_{j}^{\alpha \, 2} \cos(\omega_{j}^{\alpha }t),
\end{equation}
our kernels \(\nu_\alpha\) are  
\begin{eqnarray}
\nu_\alpha(t)=-\frac{d\mu_\alpha}{dt}, \quad \mu_\alpha(0)=\Omega_\alpha.
\end{eqnarray}
Thus, after an integration by parts, we obtain
\begin{eqnarray}
&&\dot{X}(t) - \frac{i}{\hbar}[V,X](t) =
\label{eq:altlangevinX}\\
&& \bar{U}_{y}(t) - \mu_{y}(t)\hat{Y}- 
\int_{0}^{t} \mu_{y}(t-s) \dot{Y}(s) \text{d} s, 
\nonumber\\
&&\dot{Y}(t) - \frac{i}{\hbar}[V,Y](t) =\label{eq:altlangevinY}\\
&& -\bar{U}_{x}(t) + \mu_{x}(t) \hat{X}+ 
\int_{0}^{t} \mu_{x}(t-s) \dot{X}(s) \text{d} s, 
\nonumber
\end{eqnarray}
where
\begin{eqnarray}
\hat{X}=X(0),\quad \hat{Y}=Y(0)
\end{eqnarray}
are time-independent Schr\"odinger operators. Eqs.~\eqref{eq:altlangevinX} and \eqref{eq:altlangevinY} are the operator Langevin equations for
the noncommutative plane. Unlike the standard operator
Langevin equations, they are of first order in
time derivatives, which is why the random driving term
is a velocity field rather than a Langevin force. 

{\it Correlators of the random velocity field for a
thermal bath.---}
The statistical properties of the operator-valued random 
velocity fields $\bar{U}_{x}$ and $\bar{U}_{y}$ are
described by their symmetrized expectation value. To compute 
this quantity, it is necessary to specify the state of the 
bath. A common choice is to assume that the harmonic oscillators 
are canonically distributed, at temperature $T$, with respect 
to the Hamiltonian 
\begin{align}
\begin{split}
H_B= &\sum_j \left[ \frac{1}{2m_j}p_j^{x\, 2}+
\frac{1}{2}m_{j} \omega_{j}^{x \, 2}(x_j-X)^2 \right]\\
+&\sum_j \left[ \frac{1}{2m_j}p_j^{y\, 2}+
\frac{1}{2}m_{j} \omega_{j}^{y \, 2}(y_j-Y)^2 \right],
\end{split}
\end{align}
so that the expectation values of a bath observable 
$O$ is 
\begin{equation}
    \langle O \rangle = 
\frac{\text{Tr}\{O e^{-H_{B} / k_{B}T}\}}{\mathcal{Z}_B},
\end{equation}
where $\mathcal{Z}_{B}$ is the partition function of the bath.
In particular, 
\begin{align}\label{eq:velocityexpectation}
    \frac{1}{2} &\langle U_{\alpha}(t) U_{\beta}(t') + U_{\beta}(t') U_{\alpha}(t) \rangle \\
                &= \delta_{\alpha \beta} \frac{l_{B}^{4}}{2\hbar} \sum_{j} m_{j}\omega_{j}^{\alpha \, 3} \cos[\omega_{j}^{\alpha}(t-t')] \coth \left( \frac{\hbar \omega_{j}^{\alpha}}{2k_{B} T} \right), \nonumber
\end{align}
see Appendix \ref{sec:vfieldstatistics} for a derivation
of this result.

{\it The continuum frequency limit.---}
If one would like to allow for true dissipation of energy 
by way of the bath, then it becomes necessary to let the number of 
oscillators become infinite in terms of a continuum range of 
frequencies. Let us introduce an interpolating mass function 
$m_{\alpha}(\omega)$, such that
\begin{eqnarray}
m_{j} = m_{\alpha}(\omega^{\alpha}_{j}).
\end{eqnarray}
The spectral density of the bath is
\begin{equation}
\rho_{\alpha}(\omega) = \sum_{j} \delta(\omega - \omega_{j}^{\alpha}).
\end{equation}
It is understood that \(\rho(\omega)=0\) if \(\omega\leq0\).
With the help of these definitions, all of our previous 
expressions may be rewritten in terms of integrals involving  
\(\rho\) and \(\kappa\), so that the properties of the bath 
are encoded in the spectral density \(\rho\). For example, 
the memory kernel of Eq.\,\eqref{defmemker} becomes 
\begin{eqnarray}
\mu_\alpha(t)=\frac{l_{B}^{2}}{\hbar}
\int_0^\infty d\omega\, \rho_\alpha(\omega)m_{\alpha}(\omega) \omega^{2} \cos(\omega t).
\end{eqnarray}

\section{Diffusion in the noncommutative plane}
\label{topobrown}

In this section, we will use our Langevin equation, 
Eqs.\,\eqref{eq:altlangevinX} and \eqref{eq:altlangevinY}, 
for investigating an important aspect of topological Brownian 
motion: diffusion in the noncommutative plane. According
to our results in Sec.\,\ref{alternative}, we 
can think of this phenomenon as describing the emergent 
properties of normal electronic diffusion in the situation 
in which an applied magnetic field is strong enough to project 
the Brownian motion of the electron to a fixed Landau level.  

For this section, the oscillator bath is isotropic and \(V=0\), 
so that the total system as described by the Hamiltonian of 
Eq.\,\eqref{noncomelpbath} is rotationally and translationally 
invariant. Moreover, we assume that the frequency distribution
of the bath is such that the memory kernels are  
\begin{eqnarray}
\mu_x(t)=\mu_y(t)=2 \gamma\delta(t).
\end{eqnarray}
Recalling the definition of these memory kernels, 
Eq.\,\eqref{defmemker}, we see that a particular distribution 
of frequencies and masses with this this property is
\begin{equation}\label{eq:OscillatorDistribution}
\omega^{x}_{j} = \omega^{y}_{j} = j, \quad 
m_{j} = \frac{2}{j} \frac{\gamma \hbar}{\pi l^{2}_{B}}.
\end{equation}
This statement is rather qualitative, but suffices for our
purposes, see Appendix \ref{sec:markovianLangevinEquation} for 
the justifications of these claims.

Under these conditions, our Langevin equation reduces to 
\begin{align}
\dot{X}(t) + \gamma \dot{Y}(t) &= \bar{U}_{y}(t) - 2\gamma\delta(t)\hat{Y}, \\
\dot{Y}(t) - \gamma \dot{X}(t) &= - \bar{U}_{x}(t) + 2\gamma\delta(t)\hat{X}.
\end{align}
The meaning of terms proportional to \(\delta(t)\) was 
explained in Sec.\,\ref{standardole}. We will discard these terms 
with the understanding that \(t\geq0\) always in the following.
Then, an elementary calculation yields
\begin{align}
X(t) =& \frac{1}{\gamma^{2}+1} \int_{0}^{t} \left[ \bar{U}_{y}(s) 
+ \gamma \bar{U}_{x}(s)\right] \text{d}s + \hat{X}, \label{Xdiff}\\
Y(t) =& \frac{1}{\gamma^{2}+1} \int_{0}^{t}\left[ -\bar{U}_{x}(s)  
+ \gamma \bar{U}_{y}(s)\right] \text{d}s + \hat{Y}.
\end{align}
While the parameter \(\gamma\) has the interpretation of a 
friction coefficient, the Hamiltonian of a free electron vanishes 
identically in the noncommutative plane. Hence, the electron 
has no energy to give to the bath, and cannot take energy from 
the bath either. It is a bizarre setup for diffusion, and 
further insight would be very desirable.

\begin{widetext} 
One of the hallmarks of classical Brownian motion is that its mean-squared displacement grows linearly as a function of time. Hence, let us compute 
\begin{eqnarray}
\langle R^2(t)\rangle=\lim_{t'\rightarrow t} 
\frac{1}{2} \langle X(t)X(t') + X(t')X(t) \rangle
+\frac{1}{2}\langle Y(t)Y(t') + Y(t')Y(t) \rangle,
\end{eqnarray}
the mean displacement (from the origin) squared. Since the 
system is rotationally invariant, this expression simplifies to
\begin{eqnarray}
\langle R^2(t)\rangle=
\lim_{t'\rightarrow t} \langle X(t)X(t') + X(t')X(t) \rangle=
\lim_{t'\rightarrow t}\langle Y(t)Y(t') + Y(t')Y(t) \rangle.
\end{eqnarray}
\end{widetext}

At this point, it becomes necessary to make an explicit
choice of state for the system. We take the product state
\begin{eqnarray}
\rho=\frac{e^{- H_B/k_BT}}{\mathcal{Z}_B}\otimes |0\rangle\langle 0|, 
\end{eqnarray}
where \(|0\rangle\) denotes the coherent state centered
at the origin of the noncommutative plane, see 
Sec.\,\ref{quantihigh}. Since the operator 
Langevin equation is derived in the Heisenberg picture,
this state has the following interpretation at \(t=0\): 
The electron is maximally localized at the origin, the
bath is in a thermal Gibbs state at temperature \(T \), and there is no entanglement 
between the two. Recall that, by convention, \(t=0+\) is 
the time when the Heisenberg, Schr\"odinger, and interaction 
pictures entering the operator Langevin equation coincide
(the reasons for writing \(t=0+\) are explained in 
Sec.\,\ref{standardole}).

\begin{widetext}
Our choice of state for the system implies 
\begin{eqnarray}
\langle \hat{X}\rangle=0=\langle\hat{Y}\rangle,\quad 
\langle\hat{X}^2\rangle=\frac{l_B^2}{2}=\langle\hat{Y}^2\rangle,
\quad \langle \bar{U}_{x}(t) \bar{U}_{y}(t') \rangle = 0.
\end{eqnarray} 
Combining this information with Eq.\,\eqref{Xdiff}, we obtain
\begin{align} \label{eq:symmexpectationvalue}
\frac{1}{2}\langle &X(t)X(t') + X(t')X(t) \rangle \\
&= \frac{1}{2(\gamma^{2}+1)^{2}}\int_{0}^{t} 
\text{d}s \int_{0}^{t'} \text{d}s' \bigg\langle 
\bar{U}_{y}(s) \bar{U}_{y}(s') + 
\bar{U}_{y}(s')\bar{U}_{y}(s) + 
\gamma^{2} \bar{U}_{x}(s) \bar{U}_{x}(s') + 
\gamma^{2}\bar{U}_{x}(s') \bar{U}_{x}(s) \bigg\rangle + 
\frac{l_B^2}{2}. \nonumber
\end{align}

With our current choice of distribution for the oscillator 
masses and frequencies, Eq.~\eqref{eq:velocityexpectation} 
reduces to 
\begin{align}
\frac{1}{2} \langle \bar{U}_{\alpha}(t) \bar{U}_{\beta}(t') + 
\bar{U}_{\beta}(t') \bar{U}_{\alpha}(t) \rangle 
&= \delta_{\alpha \beta} \frac{\gamma l_{B}^{2}}{\pi} \sum_{j} j^{2} \cos[j(t-t')] \coth \left( \frac{\hbar j}{2k_{B} T} \right) \nonumber \\
                                                                                          &= \delta_{\alpha \beta} \frac{\gamma l^{2}_{B}}{\pi} \int_{0}^{\infty}\omega \cos[\omega(t-t')] \coth \left( \frac{\hbar \omega}{2k_{B}T} \right)\text{d}\omega,
\end{align}
Strictly speaking, this integral diverges. However, one 
can make sense of it by declaring it to be the Fourier 
cosine transform of $\omega \coth(\hbar \omega/(2k_{B}T))$.  
This yields (see Ref.~[\onlinecite{FLO87}])
\begin{equation}
    \frac{1}{2}\langle \bar{U}_{\alpha}(t) \bar{U}_{\beta}(t') + 
\bar{U}_{\beta}(t')\bar{U}_{\alpha}(t) \rangle = 
\delta_{\alpha \beta} k_{B} T \frac{\gamma l_{B}^{2}}{\hbar} 
\der{}{t} \coth \left( \frac{\pi k_{B} T(t-t')}{\hbar} \right).
\end{equation}
The symmetrized velocity correlator does not distinguish between 
the $x$ and $y$-direction, as was to be expected, since the 
distributions for the masses and frequencies were chosen to be 
identical as well, see Eq.~\eqref{eq:OscillatorDistribution}. 
Hence, Eq.~\eqref{eq:symmexpectationvalue} becomes
\begin{equation}\label{eq:easyintegral}
\frac{1}{2} \langle X(t) X(t') + X(t')X(t) \rangle = 
\frac{\gamma k_{B} T l^{2}_{B}}{ \hbar(1+\gamma^{2})}\int_{0}^{t'} \text{d}s' \coth \left( \frac{\pi k_{B} T(t-s')}{\hbar} \right)
\end{equation}
where we have dropped the unimportant constant \(\langle \hat{X}^{2} \rangle\).
Finally, by performing the integral in Eq.~\eqref{eq:easyintegral} for 
$0 < t' <t$ we obtain
\begin{equation}
\frac{1}{2} \langle X(t)X(t') + X(t')X(t) \rangle = 
\frac{\gamma l^{2}_{B}}{\pi(1+\gamma^{2})} 
\log \left[ \text{csch}\left(\frac{\pi(t-t') T k_{B}}{\hbar}\right)\text{sinh}\left( \frac{\pi t T k_{B}}{\hbar}\right) \right].
\end{equation}
\end{widetext}

The next step in order to compute the mean squared displacement
is taking the limit \(t'\rightarrow t\). However, our current 
expression is not valid for $t=t'$. Hence, we set $t' = t - \eps$, 
with $\eps$ is some small positive time, so that 
\begin{align}
    &\frac{1}{2} \langle X(t)X(t-\eps) + X(t-\eps)X(t) \rangle\\
                    &= C(\eps) + \frac{ \gamma l^{2}_{B}}{\pi (1+\gamma^{2})}\log\left[ \sinh \left( \frac{\pi t T k_{B}}{\hbar} \right) \right].
\nonumber
\end{align}
The point to notice is that  
\begin{equation}
    C(\eps) = \frac{\gamma l_{B}^{2}}{\pi (1+ \gamma^{2})} \log \left[ \text{csch}\left( \frac{\pi \eps T k_{B}}{\hbar}\right) \right]
\end{equation}
is independent of \(t\). For large times \(t\rightarrow \infty\), 
\begin{align}
    \log&\left[ \sinh \left( \frac{\pi t T k_{B}}{\hbar} \right) \right] \nonumber \\
                     &\approx \log \left[ \frac{1}{2} \exp \left( \frac{\pi t T k_{B}}{\hbar}\right) \right] = \frac{\pi T k_{B}}{\hbar}t - \log[2],
\end{align}
which leads to
\begin{align}
    \frac{1}{2} \langle X(t)&X(t-\eps) + X(t-\eps)X(t) \rangle \nonumber \\
                            &= \frac{ \gamma l_{B}^{2} T k_{B}}{ \hbar (1+\gamma^{2})} t + C'(\eps),
\end{align}
with 
\begin{equation}
{C'}(\eps) = \frac{\gamma}{\pi (1+\gamma^{2})} \log 
\left[\frac{1}{2} \text{csch}
\left( \frac{\pi \eps T k_{B}}{\hbar}\right) \right]
\end{equation}
again independent of time. Hence, to leading order in \(t\),
\begin{equation}\label{eq:LargeBDeviation}
\langle R^2(t)\rangle=2\langle X^{2}(t) \rangle =
2 \frac{\gamma l_{B}^{2} k_{B} T}{  \hbar( 1+ \gamma^{2})}t.
\end{equation}
This is precisely the behavior characteristic of normal 
diffusion! 

In closing, we would like to compare the result 
Eq.\,\eqref{eq:LargeBDeviation} with the analogous
result for the standard quantum Langevin equation, 
Eq.\,\eqref{standarddiff}. For this equation
of motion, in the classical limit one obtains\cite{gardiner}
\begin{equation}
\langle X^{2}(t)\rangle = 2 \frac{k_{B} T}{\eta} t.
\end{equation}
Now, by dimensional analysis, 
\begin{equation}
	\gamma = \frac{l_{B}^{2}}{\hbar} \eta.
\end{equation}
Hence, we can reexpress our result, Eq.~\eqref{eq:LargeBDeviation}, as
\begin{equation}
\langle R^{2}(t) \rangle =  
\frac{2k_{B}T}{\eta + \frac{\hbar^{2}}{l_{B}^{4}} \frac{1}{\eta} }t.
\end{equation}
This shows, on one hand, that 
\begin{equation}
\langle R^{2}(t) \rangle = 2\frac{k_{B}T}{\eta} t
\end{equation}
in the limit $\eta \gg \hbar/l_{B}^{2}$, in which the 
friction coefficient becomes dominant. This is precisely
the classical result just mentioned above. On the other 
hand, if the friction constant is small, 
$0 < \eta \ll \hbar /l_{B}^{2}$, then
\begin{equation}
\langle R^{2}(t) \rangle \approx 
2\frac{l_{B}^{4} k_{B} T \eta}{\hbar^{2}}t =
2\frac{c^{2} k_{B} T \eta}{(eB)^{2}} t.
\end{equation}
This is in stark contrast with the classical result: If $\eta$ is 
small, the magnetic field suppresses fluctuations. This is a very 
important result of this work.

\section{Summary and outlook}
\label{outlook}

In this paper, we have investigated an instance of
topological quantum Brownian motion: an electron 
subjected to an ultra-high magnetic field and coupled
to a thermal bath of independent oscillators. The operator
Langevin equations that we derive for modeling this
system are unconventional. They are first-, rather than
second-order differential equations, and can
be interpreted as describing quantum Brownian motion
projected onto a Landau level. In spite of the 
differences between the standard and our Langevin equation, 
diffusion in the noncommutative plane, or equivalently,
in a Landau level, is conventional, i.e.~the mean squared
displacement is proportional to \(t\). However, the
proportionality constant displays a bizarre regime
as a function of the friction coefficient: For strong dissipation, friction reduces diffusion, as expected, and the magnetic field plays no role, but for weak dissipation, friction enhances diffusion and the magnetic field suppresses it. It would be remarkable
to observe this regime experimentally, but it is far from clear
how to do so. Possibly this is yet another problem for the 
fast growing field of quantum simulations. 

We also investigated other physical aspects of the noncommutative
plane. We explored its unconventional symmetries, analyzed the 
fact that the electron cannot be perfectly localized due
to Heisenberg uncertainty, and showed that quantum tunneling can 
be completely suppressed in the noncommutative plane. Since quantum 
tunneling out of a metastable minimum is often thought of as an unavoidable 
fact of (quantum) life, it is interesting to see a regime where it can,
in fact, be avoided. The complete suppression of quantum tunneling,
together with the suppression of Brownian fluctuations by the magnetic 
field in the weak-friction regime, suggest that the noncommutative 
plane might be well suited for novel designs of quantum memories.

Let us conclude with an open problem. 
Since the operator Langevin equation is either simple or 
impossible to solve, according to whether the potential 
\(V(X,Y)\) is at most quadratic in the coordinates, it is natural 
to attempt a path integral description of the problem. It 
is not difficult to write a suitable propagator,  
\begin{align}
&G(Q'',T;Q',0)=\\
&\int {\cal D}Q \exp \left\{ \frac{i}{\hbar}\int_0^T
dt\,\left[\frac{\hbar}{l_B^2}Y\dot{X}-V(X,Y)\right] \right\}\nonumber\\
\times &\exp \left\{ \frac{i}{\hbar}\int_0^T\sum_j \left[ \frac{1}{2}m_j\dot{x}_j^2
-\frac{1}{2}m_j\omega_{j}^{x \, 2}(x_j-X)^{2}\right] \right\} \nonumber\\
\times &\exp \left\{ \frac{i}{\hbar}\int_0^T\sum_j \left[ \frac{1}{2}m_j\dot{y}_j^2
-\frac{1}{2}m_j\omega_{j}^{y \, 2}(y_j-Y)^{2} \right] \right\}, \nonumber
\end{align}
where \(Q'=\{X',Y',x_j',y_j'\}\) and \(Q''\) 
denote collectively the initial and final values of the 
coordinates, respectively. Moreover, it is possible
to integrate out the bath as usually to obtain a description
of the electron-bath coupling in terms of influence functionals.\cite{feynman}
The problem is that the resulting effective path integral
for the electron has the structure of a phase-space 
path integral, rendering semiclassical approximations 
untrustworthy.\cite{schulman} What is then a good way of 
modeling nonlinear forces and friction in the noncommutative plane?

\acknowledgements

EC gratefully acknowledges discussions with L. Viola. This work is part 
of the DITP consortium, a program of The Netherlands Organisation for
Scientific Research (NWO) that is funded by the Dutch Ministry of 
Education, Culture and Science (OCW).

\appendix

\section{Statistical Properties of the velocity field}
\label{sec:vfieldstatistics}

In this section we recall the arguments of 
Ref.~[\onlinecite{FKM65}] leading to 
Eq.\,\eqref{eq:velocityexpectation}. Consider a system of 
decoupled harmonic oscillators, canonically distributed at 
temperature $T$ and described by the Hamiltonian
\begin{equation}\label{eq:SHOHamiltonian}
H_{B} = \sum_{j}\left[ \frac{p_{j}^{2}}{2m_{j}} + \frac{1}{2} m_{j} \omega_{j}^{2} x_{j}^{2} \right].
\end{equation}
If $O$ is any observable, then its expectation value 
may be computed as
\begin{equation}
	\langle O \rangle = \frac{ \text{Tr} \{ O \exp( - H_{B}/(kT)) \}}{\text{Tr}\{ \exp ( - H_{B} / (kT)) \}}.
\end{equation}

The usual annihilation operator
\begin{equation}
	a_{j} = \sqrt{\frac{m_{j} \omega_{j}}{2 \hbar}} 
\left( \hat{x}_{j} + \frac{i}{m_{j} \omega_{j}}\hat{p}_{j} \right)
\end{equation}
and its adjoint allows us to rewrite 
Eq.~\eqref{eq:SHOHamiltonian} as
\begin{equation}
	H_{B} = \sum_{j} \hbar \omega_{j} \left( a_{j}^{\dagger}a_{j} + \frac{1}{2} \right).
\end{equation}
The eigenvalues of the $j$-th single particle Hamiltonian 
$H_{Bj} = \hbar \omega_{j} (a_{j}^{\dagger} a_{j} + 1/2)$ 
are $E_{Bjn} = \hbar \omega_{j} (n + 1/2)$. Therefore, 
\begin{align}\label{eq:SHOExpectationStep}
	\langle a_{i}^{\dagger} a_{j} \rangle &= \delta_{ij} \frac{ \text{Tr} \{ a_{i}^{\dagger}a_{j} \exp( - H_{B}/(kT)) \}}{\text{Tr}\{ \exp ( - H_{B} / (kT)) \}} \\
	&= \delta_{ij} \frac{\sum_{n=0}^{\infty} n\exp \left[ -\frac{\hbar\omega_{j}}{k T}\left( n + \frac{1}{2} \right) \right]}{\sum_{n=0}^{\infty} \exp \left[ -\frac{\hbar\omega_{j}}{k T}\left( n + \frac{1}{2} \right) \right]} \nonumber.
\end{align}

In order to evaluate this sum we define the function
\begin{equation}
	f(\lambda) = \sum_{n=0}^{\infty} \exp \left[ -\frac{\hbar\omega_{j}}{k T}\left( \lambda n + \frac{1}{2} \right) \right],
\end{equation}
which can be evaluated using the geometric series. 
In terms of $f$, Eq.~\eqref{eq:SHOExpectationStep} can 
be written as
\begin{equation}
	\langle a_{i}^{\dagger} a_{j} \rangle = \delta_{ij} \frac{\partial_{\lambda} f(\lambda)}{f(\lambda)} \bigg|_{\lambda = 1}.
\end{equation}
The result is
\begin{equation}
	\langle a_{i}^{\dagger} a_{j} \rangle = \delta_{ij} \left[ \exp \left( \frac{\hbar \omega_{j}}{kT} \right) - 1 \right]^{-1}.
\end{equation}
In other words,
\begin{align}
	\langle a^{\dagger}_{i} a_{j} \rangle = \frac{1}{2} \delta_{ij} \left[ \coth \left( \frac{\hbar \omega_{j}}{2kT} \right) - 1 \right], \\
	\langle a_{i} a^{\dagger}_{j} \rangle = \frac{1}{2} \delta_{ij} \left[ \coth \left( \frac{\hbar \omega_{j}}{2kT} \right) + 1 \right],
\end{align}
and so 
\begin{align}
\langle \hat{x}_{i} \hat{x}_{j} \rangle &= 
\delta_{ij}\frac{\hbar}{2\omega_{j}m_{j}} \coth \left( \frac{\hbar \omega_{j}}{2kT} \right), \label{eq:qqexp} \\
\langle \hat{p}_{i} \hat{p}_{j} \rangle &= 
\delta_{ij} \frac{1}{2} \hbar \omega_{j}m_{j} \coth \left( \frac{\hbar \omega_{j}}{2kT} \right), \label{eq:ppexp} \\
\langle \hat{x}_{i} \hat{p}_{j} \rangle &= 
\frac{1}{2} i \hbar \delta_{ij}. \label{eq:qpexp}
\end{align}

Now we would like to compute
\begin{equation}
\frac{1}{2} \langle \bar{U}_{\alpha}(t) \bar{U}_{\beta}(t') 
+ \bar{U}_{\beta}(t')\bar{U}_{\alpha}(t) \rangle,
\end{equation}
where
\begin{equation}
\bar{U}_{\alpha}(t) = 
\frac{l_{B}^{2}}{\hbar} \sum_{j} 
m_{j} \omega_{j}^{\alpha \, 2} \left( \cos(\omega_{j}^{\alpha}t) 
\hat{r}_{j}^{\alpha} + 
\frac{ \sin(\omega_{j}^{\alpha}t)}{m_{j} \omega_{j}^{\alpha}}
\hat{p}_{j}^{\alpha} \right).
\end{equation}
Since harmonic oscillators in multiple dimensions can be 
seen as decoupled one-dimensional harmonic oscillators, 
generalizing Eqs.~\eqref{eq:qqexp}, \eqref{eq:ppexp} 
and \eqref{eq:qpexp} to multiple dimensions is 
straightforward. This reasoning yields the expectation values 
\begin{align*}
\langle \hat{r}^{\alpha}_{i} \hat{r}^{\beta}_{j} 
+ \hat{r}^{\beta}_{j} \hat{r}^{\alpha}_{i} \rangle &= 
\delta_{\alpha \beta} \delta_{ij}\frac{\hbar}{\omega^{\alpha}_{j}m_{j}} \coth \left( \frac{\hbar \omega_{j}}{2kT} \right), \\
\langle \hat{p}^{\alpha}_{i} \hat{p}^{\beta}_{j} + 
\hat{p}^{\beta}_{j} \hat{p}^{\alpha}_{i} \rangle &= 
\delta_{\alpha \beta} \delta_{ij} \hbar \omega^{\alpha}_{j}m_{j} \coth \left( \frac{\hbar \omega_{j}}{2kT} \right), \\
\langle \hat{r}^{\alpha}_{i} \hat{p}^{\beta}_{j} +
 \hat{p}^{\beta}_{j} \hat{r}^{\alpha}_{i} \rangle &= 0,
\end{align*}
and so we obtain
\begin{align}
	&\frac{1}{2} \langle \bar{U}_{\alpha}(t) \bar{U}_{\beta}(t') + 
\bar{U}_{\beta}(t')\bar{U}_{\alpha}(t) \rangle \nonumber \\
	=& \frac{l_{B}^{4}}{2\hbar^{2}}\delta_{\alpha \beta} \sum_{j} m_{j}^{2} \omega_{j}^{\alpha \, 4} \bigg[ \cos(\omega_{j}^{\alpha}t) \cos(\omega_{j}^{\alpha}t') 2\langle \hat{r}_{j}^{\alpha} \hat{r}_{j}^{\alpha} \rangle \nonumber \\
	& \quad \quad + \sin(\omega_{j}^{\alpha}t) \sin(\omega_{j}^{\alpha}t') 2 \frac{\langle \hat{p}_{j}^{\alpha} \hat{p}_{j}^{\alpha}\rangle}{m_{j}^{2}\omega_{j}^{\alpha \, 2}} \bigg] \nonumber \\
	=& \delta_{\alpha \beta} \frac{l_{B}^{4}}{2\hbar} \sum_{j} m_{j}\omega_{j}^{\alpha \, 3} \cos[\omega_{j}^{\alpha}(t-t')] \coth \left( \frac{\hbar \omega_{j}^{\alpha}}{2k T} \right).
\end{align}

\section{Markovian Langevin equation}
\label{sec:markovianLangevinEquation}

In this appendix we show that the distributions for the masses and the frequencies of the bath oscillators given by
\begin{equation}\label{eq:appDistribution}
	\omega_{j}^{x} = \omega_{j}^{y} = j, \quad m_{j} = \frac{2}{j} \frac{\gamma \hbar}{\pi l_{B}^{2}}
\end{equation}
lead to instantaneous memory kernels, in particular this choice gives
\begin{equation}
	\int_{0}^{t} \mu_{x}(t-s) \dot{X}(s) \text{d} s = \gamma \dot{X}(t).
\end{equation}
In the following we will omit the labels $x$ and $y$, the argument is valid for both. Substitution of Eq.~\eqref{eq:appDistribution} into Eq.~\eqref{defmemker} yields
\begin{equation}
	\mu(t) = \frac{2\gamma}{\pi} \sum_{j} \cos( j t).
\end{equation}
Now, we take the continuum limit of this expression and perform the integral
\begin{equation}
	\mu(t) \longrightarrow \frac{2 \gamma}{\pi} \int_{0}^{\infty} \cos(j t) \text{d}j = 2 \gamma \delta(t)
\end{equation}
If we now define the Heaviside $\theta$-function by
\begin{equation}
	\theta(t) := \int_{-1}^{t} \delta(s) \text{d} s,
\end{equation}
we obtain
\begin{equation*}
	\int_{0}^{t} \mu(t-s) \dot{X}(s) \text{d} s = 2 \gamma \theta(0) \dot{X}(t).
\end{equation*}
The value $\theta(0)$ depends on the limiting procedure used in defining the Dirac delta distribution. In this case, we may write
\begin{align}
\begin{aligned}
	\theta(0) &= \int_{-1}^{0} \delta(s) \text{d} s \\
	&= \frac{1}{\pi} \int_{-1}^{0} \text{d} s \int_{0}^{\infty} \text{d} k  \cos[k s] \\
	&= \frac{1}{\pi} \int_{0}^{\infty} \text{d} k \int_{-1}^{0} \cos[ks] \\
	&= \frac{1}{\pi} \int_{0}^{\infty} \text{d} k \frac{\sin[k]}{k} \\
	&= \frac{1}{2}.
\end{aligned}
\end{align}
Hence
\begin{equation}
	\int_{0}^{t} \mu(t-s) \dot{X}(s) \text{d} s = \gamma \dot{X}(t).
\end{equation}


\begin{thebibliography}{99}
\bibitem{noc}
It is however possible also to have dimensional reduction and
associated topological quantum ordering by way of symmetries.\cite{noc1}
\end{thebibliography}
\end{document}